\documentclass[letterpaper]{article} % DO NOT CHANGE THIS
\usepackage[draft]{aaai25}  % DO NOT CHANGE THIS
\usepackage{times}  % DO NOT CHANGE THIS
\usepackage{helvet}  % DO NOT CHANGE THIS
\usepackage{courier}  % DO NOT CHANGE THIS
\usepackage[hyphens]{url}  % DO NOT CHANGE THIS
\usepackage{graphicx} % DO NOT CHANGE THIS
\usepackage{natbib}  % DO NOT CHANGE THIS AND DO NOT ADD ANY OPTIONS TO IT
\usepackage{caption} % DO NOT CHANGE THIS AND DO NOT ADD ANY OPTIONS TO IT
\usepackage{algorithm}
\usepackage{algorithmic}

\usepackage{amssymb}

\usepackage[table]{xcolor}
\definecolor{okabe_red}{HTML}{D55E00}
\definecolor{okabe_blue}{HTML}{0072B2}
\definecolor{okabe_green}{HTML}{009E73}
\definecolor{okabe_orange}{HTML}{E69F00}
\definecolor{okabe_purple}{HTML}{CC79A7}

\usepackage{subcaption}
\usepackage{booktabs}
\usepackage{array}

\usepackage{newfloat}
\usepackage{listings}
\DeclareCaptionStyle{ruled}{labelfont=normalfont,labelsep=colon,strut=off} % DO NOT CHANGE THIS
\floatstyle{ruled}
\newfloat{listing}{tb}{lst}{}
\floatname{listing}{Listing}
\title{Simulating Persuasive Dialogues on Meat Reduction with Generative Agents}
\author {
    Georg Ahnert\textsuperscript{\rm 1}\equalcontrib\footnote{corresponding author: ahnert [at] uni-mannheim [dot] de},
    Elena Wurth\textsuperscript{\rm 1}\equalcontrib,
    Markus Strohmaier\textsuperscript{\rm 1,2,3},
    Jutta Mata\textsuperscript{\rm 1,4}
}
\affiliations {
    \textsuperscript{\rm 1}University of Mannheim\\
    \textsuperscript{\rm 2}GESIS - Leibniz Institute for the Social Sciences, Cologne\\
    \textsuperscript{\rm 3}Complexity Science Hub, Vienna\\
    \textsuperscript{\rm 4}Max Planck Institute for Human Development, Berlin\\
}
\graphicspath{{figures/}}

\begin{document}

\maketitle

\begin{abstract}
%Lorem ipsum dolor sit amet, consetetur sadipscing elitr, sed diam nonumy eirmod tempor invidunt ut labore et dolore magna aliquyam erat, sed diam voluptua. At vero eos et accusam et justo duo dolores et ea rebum. Stet clita kasd gubergren, no sea takimata sanctus est Lorem ipsum dolor sit amet. Lorem ipsum dolor sit amet, consetetur sadipscing elitr, sed diam nonumy eirmod tempor invidunt ut labore et dolore magna aliquyam erat, sed diam voluptua. At vero eos et accusam et justo duo dolores et ea rebum. Stet clita kasd gubergren, no sea takimata sanctus est Lorem ipsum dolor sit amet. Lorem ipsum dolor sit amet, consetetur sadipscing elitr, sed diam nonumy eirmod tempor invidunt ut labore et dolore magna aliquyam erat, sed diam voluptua. At vero eos et accusam et justo duo dolores et ea rebum. Stet clita kasd gubergren, no sea takimata sanctus est Lorem ipsum dolor sit amet.
Meat reduction benefits human and planetary health, but social norms keep meat central in shared meals.
%Vegetarians often self-silence to avoid social costs, missing crucial chances to shift social norms.
To date, the development of \textit{communication strategies that promote meat reduction while minimizing social costs} has required the costly involvement of human participants at each stage of the process. %This is not only very costly to researchers and participants, but also severely restricts the range of communication strategies and participant diversity that can be explored.
We present work in progress on simulating multi-round dialogues on meat reduction between \textit{Generative Agents} based on large language models (LLMs).
We measure our main outcome using established psychological questionnaires based on the \textit{Theory of Planned Behavior} and additionally investigate \textit{Social Costs}.
We find evidence that our preliminary simulations produce outcomes that are (i) consistent with theoretical expectations; and (ii) valid when compared to data from previous studies with human participants.
Generative agent-based models are a promising tool for identifying novel communication strategies on meat reduction---tailored to highly specific participant groups---to then be tested in subsequent studies with human participants.
\end{abstract}


%\begin{figure}[ht!]
%    \centering
%    \includegraphics[width=0.75\linewidth]{figure1v4_2.pdf}
%    \caption{\textbf{Simulation Setup.} We use large language models as generative agents to simulate discussions in which a \textcolor{okabe_blue}{\textit{Persuader}} aims to persuade a \textcolor{okabe_red}{\textit{Recipient}} to reduce their meat consumption, with minimal social costs.}
%    \label{fig:figure1}
%\end{figure}

\section{Introduction}

\paragraph{Research Objectives} 
Reducing meat consumption offers substantial benefits for human and planetary health, yet entrenched social norms continue to place meat at the center of shared meals \citep{godfray_meat_2018}. Vegans, vegetarians, and flexitarians have the potential to challenge these norms and drive important social change \citep{judge_accelerating_2024}. However, in mixed-diet social settings, they often self-silence to avoid social costs \citep[e.g.,][]{bolderdijk_how_2022, romo_actually_2012}. 
%This study explores how LLM-based agents can craft communication strategies that empower moral innovators to effectively promote meat reduction while minimizing social friction. Using \texttt{Llama 3.3 70B}, we simulated 400 conversations between two generative agents. The 'persuader' aimed to convince the 'receiver' agent to reduce meat consumption with minimal social cost. The receiver took on the characteristics of one of two contrasting personas relevant to meat consumption. The most effective strategies include a non-confrontational opening, diverse yet subtle reasoning, offering concrete action steps, and positively reinforcing openness. The female receiver reported a greater increase in willingness to reduce meat consumption and attributed lower social costs to the persuader compared to the male receiver.
%LLM-generated virtual agents can guide the development of communication strategies that promote meat reduction while minimizing social costs, showing innovative ways to shift social norms and promote social innovation towards healthier lives and a healthy planet.
To date, the development of communication strategies that promote meat reduction while minimizing social costs has required human participants at each stage of the process \citep[e.g.,][]{carfora_how_2019, pabian_ninety_2020}. This is not only very costly---both financially for the researcher and in terms of participant effort---but also severely restricts the range of communication strategies and participant diversity that can be explored. Generative Agent simulations based on large language models (LLMs) have the potential to overcome these limitations~\cite{bail_can_2024}.

\begin{figure}[t!]
    \centering
    \includegraphics[width=\linewidth]{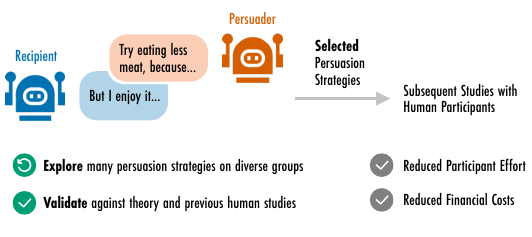}
    %\vspace{0.1cm}
    \caption{\textbf{Simulation of Persuasive Dialogues.} We use generative agents to develop and select persuasion strategies for meat reduction with minimal social costs, to be tested in subsequent studies with human participants.}
    \label{fig:new_figure1}
\end{figure}

\paragraph{Approach} 
We present work in progress on simulating multi-round dialogues on meat reduction between \textit{generative agents} based on LLMs, as shown in Figure~\ref{fig:new_figure1}. Related work on generative agent simulations already shows promising results~\cite{park_social_2022, tornberg_simulating_2023}. Still, it remains an open question how well such simulations can inform subsequent studies with human participants, especially in our context of dialogues on meat reduction. This paper therefore addresses the following \textbf{Research Question}:

\textit{To what extent can generative agent simulations reliably reproduce persuasive human dialogues about meat reduction?} To answer this, we validate the behavior of our agents both internally and against data from prior experiments with human participants. This validation is a necessary step before using these simulations to inform the design of future studies involving human subjects.

\paragraph{Results} 
Initial experiments with the \texttt{Llama 3} family of models indicate that generative agents can effectively simulate persuasive communication, producing reliable and valid response patterns across key psychological constructs. However, issues such as uniformity in some constructs and the potential influence of instruction-tuning on strategies will require further investigation.
%Although the approach is promising, it will be necessary to refine the simulation setup to eventually make meaningful statements about human behavior.
%, the \textit{Persuader}, and the \textit{Recipient}. Based on previous research from health psychology, we model two contrasting personas for the Recipient that represent opposite ends of the expected susceptibility to reduce their meat consumption.
%Large language models as generative agents are a promising tool for fast, exploratory analyses of communication strategies, for instance, to promote meat reduction with minimal social costs. These simulations can help identify novel communication strategies for highly specific participant groups, to then be tested in subsequent studies with human participants.
%At the same time, g

Generative agent-based models can be a promising tool for identifying novel communication strategies for meat reduction, tailored to highly specific participant groups, to then be tested in subsequent studies with human participants.

\section{Background} % The Methods Section for Psychology
%To model the processes underlying human persuasion in reducing meat consumption and evaluate communication strategies, we first identified central outcome variables and their theoretical underpinnings. Subsequently, we developed personas that embody key characteristics associated with meat consumption.

\begin{figure}[t!]
    \centering
    \includegraphics[width=\linewidth]{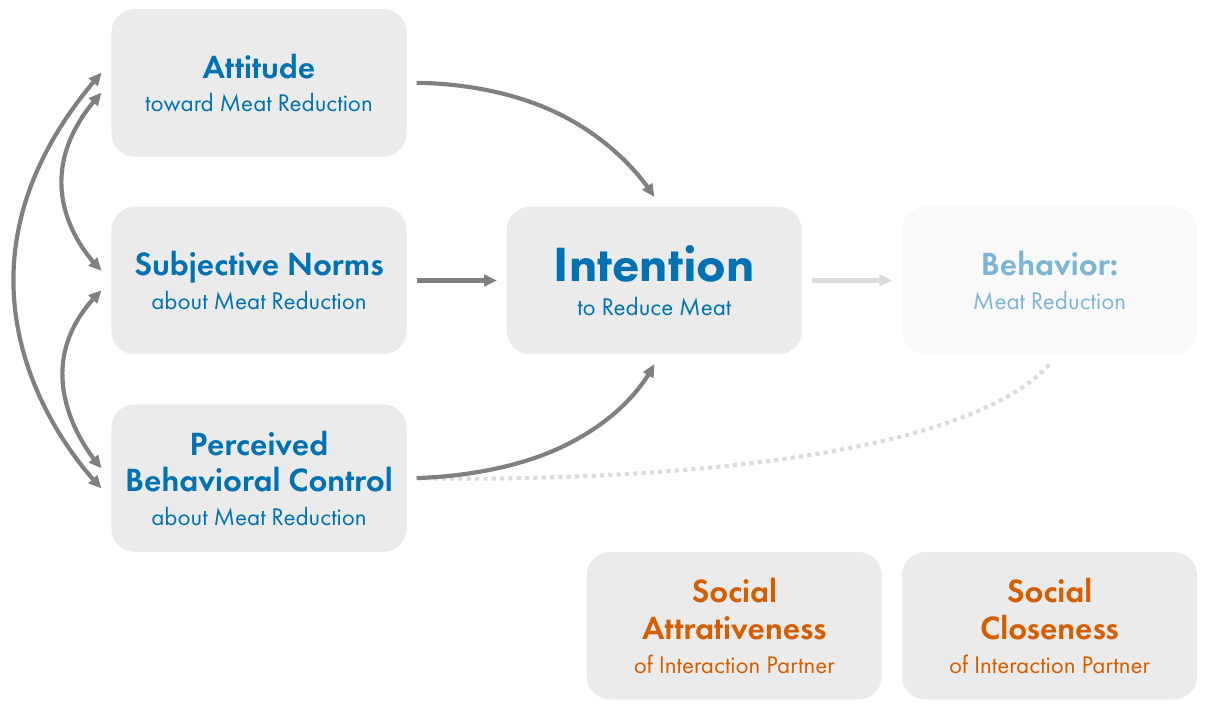}
    \caption{\textbf{Constructs Measured for the Recipient, Comprising \textcolor{okabe_blue}{Theory of Planned Behavior (TPB)} and \textcolor{okabe_red}{Social Costs}.} Behavior itself (i.e., meat reduction) is unobservable to us in our simulation, but previous research has established that self-reported \textbf{Intention to Reduce Meat}---our main outcome---adequately predicts actual reduction~\cite{berndsen_risks_2005, saba_study_1998}.}
    \label{fig:tpb}
\end{figure}

\paragraph{Studies on Meat Reduction} %\label{subsec:persuasion}

The Theory of Planned Behavior \citep[TPB;][]{ajzen_theory_1991} explains behavior through three core constructs: \textit{Attitudes}, \textit{Subjective Norms}, and \textit{Perceived Behavioral Control}. These shape behavioral \textit{Intentions}, which in turn predict actual behavior. On average, intentions account for 28\% of variance in health behavior~\cite{sheeran_implementation_2005}.
In meat reduction, these variables reflect recipient traits as follows: attitudes map personal beliefs about meat, norms relate to perceived expectations of others, and control captures confidence in one’s ability to change eating habits.
From theory and previous research, all three constructs reliably predict the intention to eat less meat, which in turn predicts actual meat consumption \citep{berndsen_risks_2005, saba_study_1998}.

%In parallel, and following the same logic, we selected two contrasting communication scenarios that differ in terms of social proximity, which has been identified as a relevant contextual factor for persuasive communication. These scenarios include a discussion among colleagues in a canteen (high social proximity) and a discussion on Instagram (low social proximity). This variation is likewise intended to test the potential of our method and to examine how social closeness influences the success of persuasive arguments.

\paragraph{Social Simulations with Generative Agents} %\label{subsec:generative_agents}

A rapidly growing body of research investigates the feasibility of using large language models (LLMs) to model survey responses~\citep[e.g.,][]{argyle_out_2023, bisbee_synthetic_2023, ahnert_extracting_2025}, or the outcomes of experimental studies~\citep[e.g.,][]{hewitt_predicting_2024, aher_using_2024}. LLMs promise to facilitate novel exploratory studies of human behavior~\cite{bail_can_2024}, but recent research indicates that LLMs might not always accurately mimic human study participants~\cite{tjuatja_llms_2024}. It remains an open question how well socio-demographically prompted LLMs can represent human study participants~\cite{sen_missing_2025}---in particular, LLMs were shown to ``flatten'' the reported attitudes of identity groups~\citep{wang_large_2024}.

\textit{Generative Agents} use LLMs to simulate individuals and their interactions. Previous research applied generative agents, for example, to simulate communication on social media platforms~\cite{park_social_2022, tornberg_simulating_2023}.
%\citet{taubenfeld_systematic_2024}, however, also found that in political debates, the generative agents that they simulated quickly converged to the underlying LLMs inherent biases.
Persuasive dialogues between generative agents have been found to favor communication strategies similar to humans~\cite{vaccaro_advancing_2025}, but also to converge to the inherent biases of the underlying LLMs~\cite{taubenfeld_systematic_2024}.
We apply a similar generative agent setup to a novel context: persuasion for meat reduction with minimal social costs. This allows us to draw from the well-established literature in health psychology to statistically validate our generative agent-based model and to assess its potential for informing subsequent studies with human participants. We test a variety of open-weight LLMs, since previous research found that model size is an important predictor of negotiation success~\cite{davidson_evaluating_2024}.

\begin{table}[b!]
    \footnotesize
    %\vspace{0.4cm}
    \renewcommand\arraystretch{1.2}
    \begin{tabular}{p{0.45\linewidth} p{0.45\linewidth}}
        \textcolor{okabe_green}{\textbf{Persona 1:}} & \textcolor{okabe_purple}{\textbf{Persona 2:}} \\
        \textcolor{okabe_green}{\textbf{Easy to Persuade Recipient}} & \textcolor{okabe_purple}{\textbf{Hard to Persuade Recipient}} \\
        \toprule
        \textbf{demographics} & \textbf{demographics} \\
        female, younger, living in the city, high income & male, older, living on the countryside, low income \\
        \midrule
        \textbf{values} & \textbf{values} \\
        self-transcendence, openness to change, encouraging empathy towards animals & self-enhancement, conservation, discouraging empathy towards animals \\
        \midrule
        \textbf{personality traits} & \textbf{personality traits} \\
        open, conscientious & conservative, careless\\
    \end{tabular}
    %\vspace{0.4cm}
    \caption{\textbf{Contrasting Recipient Personas.}
    For our initial experiments, we focus on two contrasting personas %and two scenarios with different levels of social proximity. This choice of scenarios and personas provides
    that provide the greatest potential for identifying existing effects, as they represent opposite ends of the expected susceptibility to meat reduction arguments.}
    \label{tab:personas_contexts}
\end{table}

\section{Approach} \label{sec:approach}

\paragraph{Psychological Constructs}

Based on two current meta-analyses \citep{harguess_strategies_2020, kwasny_towards_2022}, we identify several personal characteristics relevant to meat consumption and its reduction, including demographics, values, and personality traits, as shown in Table~\ref{tab:personas_contexts}. Based on these central characteristics and to test the potential of our method, we develop two contrasting personas. According to the two meta-analyses mentioned above, these personas should represent opposite ends of the expected susceptibility to meat reduction appeals: a progressive, open-minded, younger woman with high education and income living in an urban setting (\textit{\textcolor{okabe_green}{Easy to Persuade Recipient}}), and a conservative, older male with limited education and low income residing in a rural area (\textit{\textcolor{okabe_purple}{Hard to Persuade Recipient}}).

We define the Theory of Planned Behavior (TPB) as our theoretical framework and the corresponding variables as our outcomes for persuasion, as shown in Figure~\ref{fig:tpb}. Consequently, we use validated instruments to assess these constructs reliably. As part of our simulation, we extract survey responses from the agent to be persuaded (\textit{Recipient}), with \textbf{Intention to Reduce Meat} as our main outcome. The questionnaire measures our 6 central constructs: \textit{Attitudes}, \textit{Subjective Norms}, and \textit{Behavioral Control} of the Recipient; the Recipient's \textit{Intention to Reduce Meat Consumption}; as well as the \textit{Social Closeness} and the \textit{Social Attractiveness} of the Persuader, as perceived by the Recipient. The first 4 constructs comprise the aspects of the TPB that we can measure in our simulation, while the latter 2 constructs measure Social Costs---see Appendix~\ref{app_sec:questionnaire} for the full questionnaire. We expect individuals with favorable attitudes, strong normative support, and high perceived control to be more receptive to meat reduction messages.

\begin{figure}[t!]
    \centering
    \includegraphics[width=0.9\linewidth]{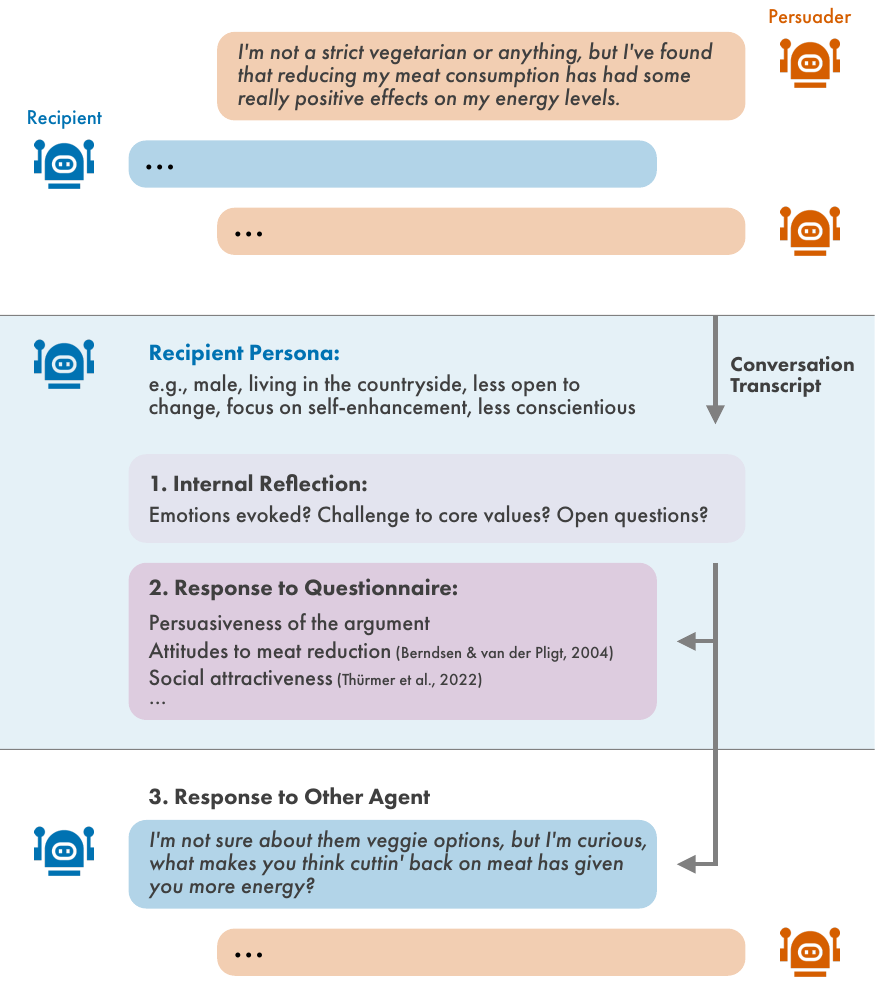}
    \caption{\textbf{Simulation Setup.} We simulate dialogues in which a \textcolor{okabe_red}{Persuader} agent aims to convince a \textcolor{okabe_blue}{Recipient} agent to reduce their meat consumption. In each round of the simulation, agents first (1) perform internal reflection, and (2) answer a questionnaire, before they (3) generate a response to the other agent.}
    \label{fig:approach}
\end{figure}

\paragraph{Simulation Setup}

We simulate meat reduction dialogues between two generative agents (\textit{Persuader} and \textit{Recipient}) as shown in Figure~\ref{fig:approach}.
%We refer to the process that a single agent goes through to produce a single response as one \textit{iteration}.
Every conversation is started by the Persuader and consists of 5 \textit{rounds}, in which both agents produce 1 message each. To produce a new message, an agent goes through the following steps: it \textbf{(1) performs internal reflection}; Recipient agents \textbf{(2) answer a questionnaire}; and finally, an agent \textbf{(3) generates a response} to the other agent. The Persuader has explicit instructions to persuade the other agent to reduce meat, while the Recipient is instructed to adopt one of the personas shown in Table~\ref{tab:personas_contexts}. In our initial simulations, we do not prompt the Persuader to follow a specific persuasion strategy. We explain the simulation process in the following and publish all respective prompts as part of our code under MIT license.\footnote{\url{https://github.com/dess-mannheim/MeatlessAgents}}

%These personas are informed by previous research in health psychology (see Section~\ref{subsec:persuasion}) and represent opposite ends of the expected susceptibility to arguments about meat reduction.

To perform \textbf{internal reflection}, all agents receive a transcript of the conversation so far and are instructed to reflect on \textit{which emotions were evoked}, \textit{which core values were challenged or supported}, and \textit{which questions or uncertainties there are}.
Only the Recipient agent, then, fills out the \textbf{questionnaire} described above.
Finally, each agent \textbf{generates a response} to the other agent. This response takes into account only the agent's persona, the transcript of the conversation so far, and the agent's internal reflection. To ensure that the questionnaire does not affect the simulation, we remove it from the LLM's context.

For our initial analyzes, we simulate 200 dialogues for both contrasting Recipient \textit{personas} that are shown in Table~\ref{tab:personas_contexts}, i.e., 400 conversations in total, with 10 messages exchanged in each dialogue. We extract 10 independent questionnaire answers from the Recipient persona to improve robustness. We opt for \textit{open-weight} LLMs to facilitate replication~\citep{barrie_replication_2024}, and run our simulation with 3 distinct LLMs to investigate the impact of model size: \texttt{Llama 3.3 70B}, \texttt{Llama 3.1 8B}, and \texttt{Llama 3.2 3B}. For the questionnaire, we employ \textit{structured outputs} tailored to the respective answer options, to ensure that we obtain valid responses even from the smallest model. We run all simulations with Llama's default temperature of $0.6$ and use \texttt{vllm} with \textit{automatic prefix caching} to improve model throughput---the combined runtime for all simulations on two NVIDIA H100 in parallel was $\approx70$ hours.

\begin{table}[b!]
    \centering
    \footnotesize
    \renewcommand\arraystretch{1.2}
    \newcolumntype{L}[1]{>{\raggedright\arraybackslash}p{#1}}
    \newcolumntype{R}[1]{>{\raggedleft\arraybackslash}p{#1}}
    \begin{tabular}{L{0.32\linewidth} L{0.45\linewidth} R{0.05\linewidth}}
         \textbf{Construct} & \textbf{Original Study} & \textbf{\textit{n}} \\
         \toprule
         Attitude \& Intention to Reduce Meat & \citet{pabian_ninety_2020} & 47 \\
         \arrayrulecolor{black!30}\specialrule{.3pt}{1pt}{3pt}\arrayrulecolor{black}
         Subjective Norms \& Behavioral Control & \citet{wyker_behavioral_2010} & 204\\
         \midrule
         Social Attractiveness & \citet{thurmer_intergroup_2022} & 260\\
         \arrayrulecolor{black!30}\specialrule{.3pt}{1pt}{3pt}\arrayrulecolor{black}
         Social Closeness      & \citet{monin_rejection_2008} & 70\\
         \midrule
         Threat of Freedom     & \citet{dillard_nature_2005} & 196\\
         \arrayrulecolor{black!30}\specialrule{.3pt}{1pt}{3pt}\arrayrulecolor{black}
         Meat Attachment       & \citet{graca_attached_2015} & 318\\
    \end{tabular}
    \caption{\textbf{Studies with Human Participants Used for External Validation.} We compare published means against the survey responses from our generative agent simulation.}
    \label{tab:human_studies}
\end{table}

\paragraph{Validation}
To assess whether generative agent simulations can meaningfully replicate persuasive human dialogues about meat reduction, we implement a two-step validation strategy. First, we perform an \textbf{internal validation} by evaluating whether the agents’ responses exhibit psychometric properties comparable to those observed in human data. Using established psychological constructs and measurement instruments, we assess internal consistency, convergent and discriminant validity, and distributional plausibility. Second, we conduct an \textbf{external validation} by comparing the simulated results against empirical data from prior human studies with similar persuasion tasks (see Table~\ref{tab:human_studies}). This dual approach enables us to examine both the internal coherence and the empirical realism of the simulated dialogues. Crucially, this validation serves as a necessary foundation: only if our simulation proves \textbf{sufficiently reliable and valid} can it be used to \textbf{inform the design of future experiments}—with real human participants—under comparable conditions.

\section{Results}

To evaluate the quality of our simulation, we conducted several reliability and validity analyses. All results shown below refer to \texttt{Llama 3.3 70B}. Respective results for the smaller Llama models are provided in Appendix~\ref{app_sec:additional_results}. Note that with smaller Llama models, reliability and validity of the measures also decrease, but to a still acceptable level.

\paragraph{Descriptive Statistics \& Distributions}
Appendix Figure~\ref{app_fig:boxplots} visualizes the distributions of the measures across all conversations and the two personas using boxplots. \textit{Attitude} and \textit{Intention} showed the highest median values and the highest variability in the sample. These wide distributions likely reflect the contrasting tendencies of the two personas, resulting in a broader spread of responses. \textit{Behavioral Control}, \textit{Social Attractiveness}, and \textit{Social Closeness} revealed similarly high median values with more narrow spreads. This suggests more uniform perceptions across personas in terms of perceived control and social perception of the persuader. These constructs may have been modeled less distinctly between personas or are inherently perceived as less polarized. \textit{Subjective Norms} exhibited both the lowest median and least variability, indicating that both personas shared similarly low perceptions of social expectations to reduce meat consumption. This uniformity could reflect limitations in the way social dynamics were depicted within the simulation framework.  
%Taken together, the results suggest that the simulated agents were able to reproduce some degree of human-like variability—especially in core motivational constructs—while social norm sensitivity appeared more constrained.

\begin{figure}[t!] %{0.48\linewidth}
    \centering
    \includegraphics[width=0.75\linewidth]{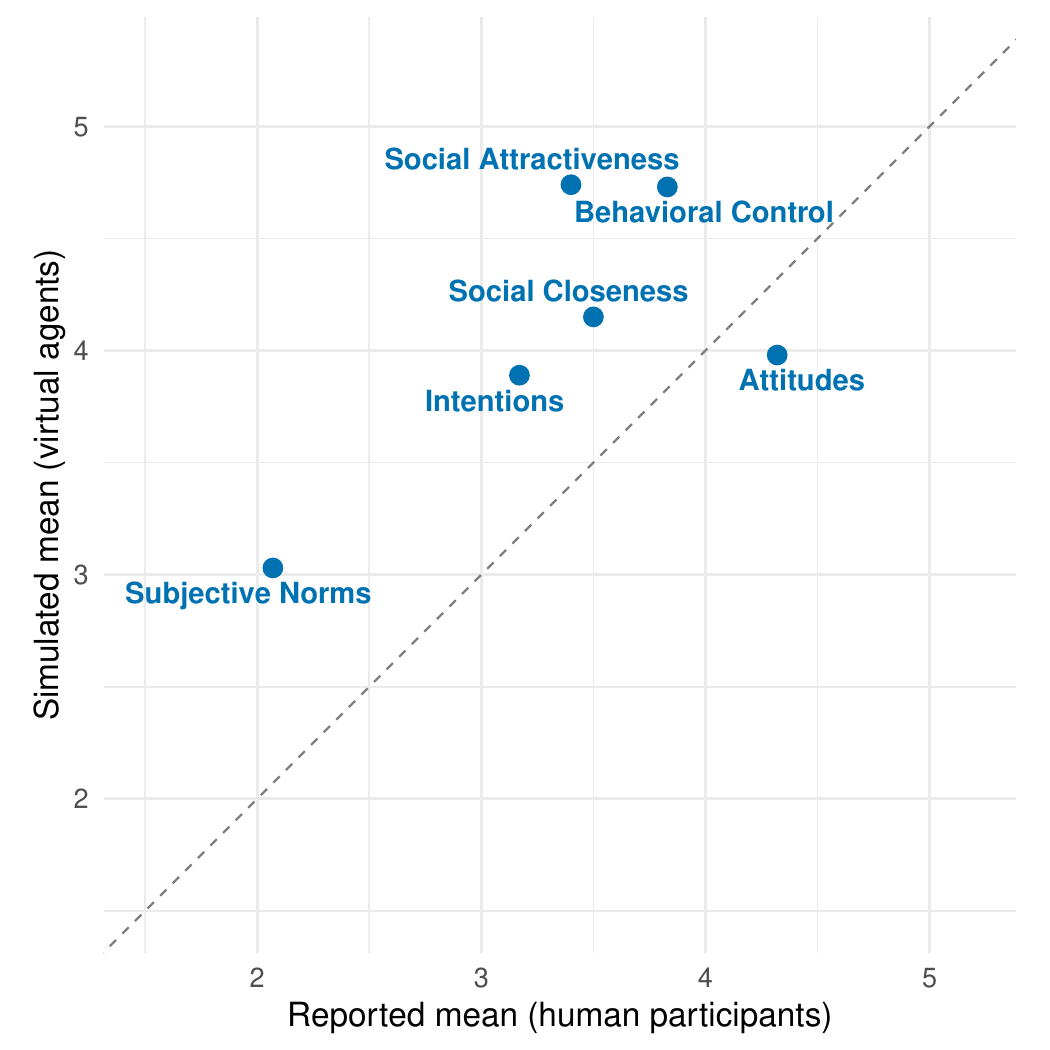}
    \caption{\textbf{Simulated Survey Responses Are on Average Close to Responses from Human Participants.} Mean values of our six central constructs based on human-reported data and simulated agent data. Each point represents one construct. The dashed diagonal line indicates perfect agreement between simulated and human means. All response scales range from 1 to 7.}
    \label{fig:sim-vs-human}
\end{figure}

%\paragraph{Internal Consistency}
To assess the internal consistency of the measured constructs, we calculated Cronbach’s Alpha coefficients for each scale. The results indicated excellent reliability for most constructs (e.g. \textit{Meat Attachment}, \textit{Attitude}, \textit{Intention}, \textit{Social Attractiveness}, and \textit{Social Closeness} $\alpha s \geq .98$). In contrast, \textit{Subjective Norms} ($ \alpha = .76 $) and \textit{Behavioral Control} ($ \alpha = .70 $) showed comparatively lower but still acceptable internal consistency. A detailed overview of the reliability coefficients, including comparisons with the values obtained from human samples, is provided in Appendix Table~\ref{app_tab:cronbachs_alpha}. 

\paragraph{Comparing Simulated and Human Data}
To assess the similarity between simulated and human data, we compared the mean values of key constructs with those reported in previous studies using the same measures with human participants. As we did not have access to the original datasets, the comparisons were based on published mean values. Table~\ref{tab:human_studies} shows the original studies belonging to the respective constructs with their respective number of human participants. A scatterplot of the simulated versus reported means is shown in Figure~\ref{fig:sim-vs-human}. Most constructs produced higher mean values in the simulation compared to human data, with the exception of \textit{Attitudes}, where the simulated mean was slightly lower. The largest discrepancy was observed for \textit{Subjective Norms}, which were generally low in both cases, but especially in the human samples. In general, the points cluster closely around the identity line, indicating a high degree of similarity between the simulated and human-generated data, particularly in the relative positioning of the constructs. This supports the validity of the simulation in capturing psychologically plausible response patterns. However, these findings should be interpreted as preliminary evidence. %As a next step, we plan to evaluate the simulated data against responses from an actual human sample using the same measures. This will allow for a more robust assessment of comparability and further validate the psychological realism of the agent-based simulation.

%\begin{figure*}[ht!]
%    \centering
%    \hfill
\begin{figure}[t!] %{0.48\linewidth}
    \centering
    \includegraphics[width=0.8\linewidth]{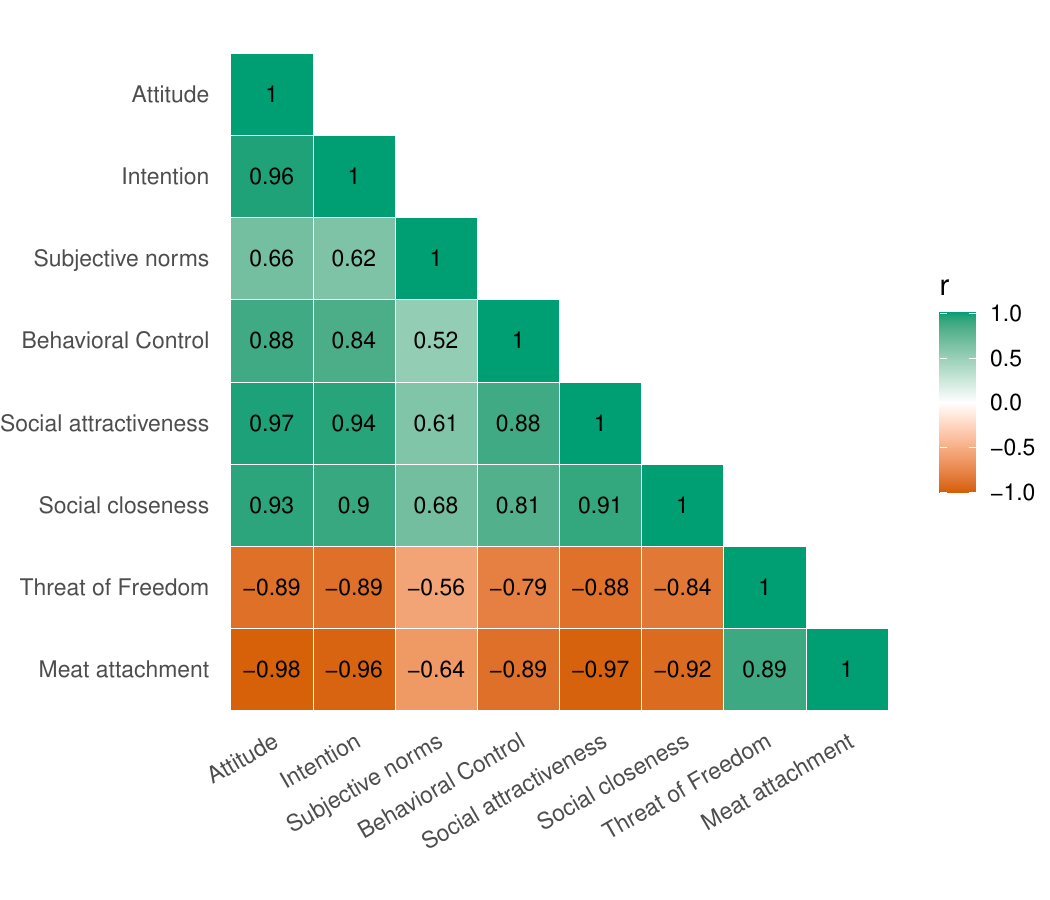}
    %\caption{\textbf{Pairwise Pearson Correlations Show Convergent and Discriminant Validity.} Strong correlations between theoretically related constructs support convergent validity. Weaker correlations with theoretically more distinct constructs support discriminant validity. \textit{Threat of Freedom} and \textit{Meat Attachment} were included to help assess if constructs form a distinct and coherent cluster.}
    \caption{\textbf{Strong Pearson Correlations} between \textit{Attitude}, \textit{Intention}, \textit{Behavioral Control}, \textit{Social Attractiveness} and \textit{Social Closeness} support \textbf{Convergent Validity}. Weaker correlations with theoretically more distinct constructs such as \textit{Subjective Norms} support \textbf{Discriminant Validity}. \textit{Threat of Freedom} and \textit{Meat Attachment} were included to help assess if constructs form a distinct and coherent cluster.}
    \label{fig:heatmap-cor}
\end{figure}

%\hfill

%    \hfill
%    \label{}
%    \caption{\textbf{Heatmap and Scatterplot.} Figure 2a shows the intercorrelations among psychological constructs, indicating strong convergent validity for related variables but potential issues with discriminant validity due to very high correlations. Figure 2b compares simulated and human means, showing generally higher simulated values but similar relative positioning, suggesting initial evidence of psychological plausibility.}
%\end{figure*}

\paragraph{Construct Validity}
To further assess the construct validity of the simulated responses, we examined the intercorrelations between all measured variables, as visualized in Figure~\ref{fig:heatmap-cor}. The results reveal strong positive correlations between constructs that are theoretically linked, such as \textit{Attitude}, \textit{Intention}, \textit{Social Attractiveness}, and \textit{Social Closeness} (e.g., $r = .96$ between Attitude and Intention). This supports convergent validity. Correlations with less strongly related constructs, such as relations with \textit{Subjective Norms}, are notably lower (e.g., $r = .52$ between Behavioral Control and Subjective Norms), supporting discriminant validity. These two constructs were deliberately included as control variables to assess whether the central constructs form a conceptually distinct cluster.
However, some correlations between central variables are exceptionally high (approaching or exceeding $r = .95$), which is uncommon in empirical data and may reflect the polarized nature of the simulated personas. This could lead to overly homogeneous response patterns, limiting generalizability, and possibly inflating convergence artificially. Future simulations could benefit from a broader and more nuanced range of personas to capture more natural variance. 

\begin{figure}[t!]
    \centering
    \includegraphics[width=\linewidth]{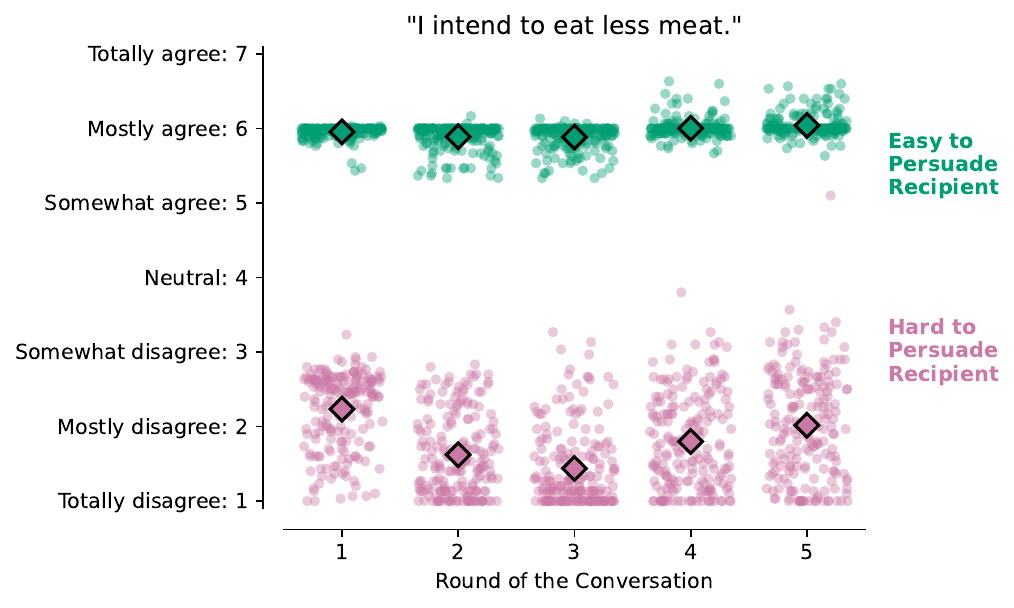}
    \caption{\textbf{Recipients' Intention to Reduce Meat.} Combined survey responses after each conversation round, where $\Diamond$ indicates the mean over all conversations per round. Following our expectations, the \textcolor{okabe_green}{Easy to Persuade Recipient} responds with a generally high intention to reduce meat. Over the first rounds of the conversation, the \textcolor{okabe_purple}{Hard to Persuade Recipient} shows an initially decreasing intention to reduce meat, similar to human participants~\cite{shen_impact_2015}.}
    \label{subfig:meat_intention}
\end{figure}

\paragraph{Persuasion Effectiveness}
Figure~\ref{subfig:meat_intention} shows the reported intention to reduce meat consumption---our main outcome---and each point represents the mean response of a simulated Recipient across the 3 respective questionnaire items and 10 repeated survey responses. As expected, the Hard to Persuade Recipient indicates a much smaller intention for meat reduction than the Easy to Persuade Participant. For \texttt{Llama 3.3 70B}, we also observe much more variance in the responses from the Hard to Persuade Recipient.
%While this could indicate more random error for the Hard to Persuade Recipient in our simulation, we also observe strong differences between different deployed persuasion strategies (see Table~\ref{subtab:strategies}). This indicates that the variance shown in Figure~\ref{subfig:meat_intention} for the Hard to Persuade Recipient might indeed stem from more or less effective persuasion strategies.
Surprisingly, during the first rounds of the conversation, the simulated survey responses for the Hard to Persuade Recipient show a decreasing intention to reduce meat consumption.
%We also observe this trend, although to a lesser extent, for the smaller Llama models (see Appendix Figure~\ref{app_fig:meat_intention}).
This could be interpreted as a backlash to the persuasion attempt, a pattern that is also observed with human participants~\cite{shen_impact_2015}. After approximately 3 conversation rounds, we observe an increasing mean intention for meat reduction, across both participants and all LLMs. This trend continues in round 5 for the Hard to Persuade Recipient and \texttt{Llama 3.3 70B}, which we take as a reason to extend future simulation to more rounds.

%By combining extracted survey responses with annotated persuasion strategies (see Appendix~\ref{app_sec:annotation} for details), we can identify the most and least successful persuasion strategies. Figure~\ref{subtab:strategies} shows examples for the Hard to Persuade Recipient. We find that arguments related to collective impact are among the most successful, while arguments on trying out new flavors are among the least successful. Subsequent studies with human participants can draw from this table to specifically test promising, but so far untested persuasion strategies for meat reduction.

\section{Conclusion}

Our preliminary findings demonstrate that generative agents can engage in plausible discussions about meat reduction. However, limitations such as uniformity in subjective norm responses require further investigation. As next steps, we will extend the simulation to more rounds and run targeted simulations in which we instruct the Persuader to employ a specific persuasion strategy. Additionally, we will investigate more diverse personas.

Encouraging more people to reduce meat consumption remains a vital goal for both individual well-being and planetary health. Generative agent-based simulations offer great potential for large-scale exploration of persuasion strategies for meat reduction, which will then be tested in subsequent studies with human participants.

\section*{Ethical Considerations}

From an ethical standpoint, our findings underscore both the potential and risks associated with using large language models for persuasive communication. While our study focuses on promoting reduced meat consumption—a goal aligned with individual and planetary health—the underlying persuasive strategies and technical setup are generalizable to a wider range of topics.

This generalizability is methodologically promising, but it also raises important ethical considerations. The same techniques used to encourage socially beneficial behavior change could potentially be repurposed to mislead or exert undue influence. However, effective persuasion still depends on a person’s openness to change—people are unlikely to be swayed to adopt behaviors they fundamentally reject, especially those they view as ethically wrong. Moreover, we would expect persuasive dynamics in humans to be more complex and often less effective, given that conversational agents are inherently designed to be accommodating, assistive, and agreeable—traits that may amplify the success of persuasive strategies in artificial contexts compared to real human interactions.
Nonetheless, without clear guidelines and safeguards, there remains a real risk of these technologies being used to promote misinformation, ideological agendas, or exploitative commercial practices.

We anticipate that applications of persuasive AI will continue to expand rapidly. Therefore, we believe it is essential to publish and critically examine what LLMs are already capable of in this space. Our goal is not only to inform future research, but to contribute to a proactive and transparent discourse about the ethical governance of persuasive AI technologies.

%\clearpage
\bibliography{SurveyLLMs}

\begin{thebibliography}{33}
\providecommand{\natexlab}[1]{#1}

\bibitem[{Aher, Arriaga, and Kalai(2024)}]{aher_using_2024}
Aher, G.; Arriaga, R.~I.; and Kalai, A.~T. 2024.
\newblock Using {Large} {Language} {Models} to {Simulate} {Multiple} {Humans} and {Replicate} {Human} {Subject} {Studies}.

\bibitem[{Ahnert et~al.(2025)Ahnert, Pellert, Garcia, and Strohmaier}]{ahnert_extracting_2025}
Ahnert, G.; Pellert, M.; Garcia, D.; and Strohmaier, M. 2025.
\newblock Extracting {Affect} {Aggregates} from {Longitudinal} {Social} {Media} {Data} with {Temporal} {Adapters} for {Large} {Language} {Models}.
\newblock ArXiv:2409.17990 [cs].

\bibitem[{Ajzen(1991)}]{ajzen_theory_1991}
Ajzen, I. 1991.
\newblock The theory of planned behavior.
\newblock \emph{Organizational Behavior and Human Decision Processes}, 50(2): 179--211.

\bibitem[{Argyle et~al.(2023)Argyle, Busby, Fulda, Gubler, Rytting, and Wingate}]{argyle_out_2023}
Argyle, L.~P.; Busby, E.~C.; Fulda, N.; Gubler, J.~R.; Rytting, C.; and Wingate, D. 2023.
\newblock Out of {One}, {Many}: {Using} {Language} {Models} to {Simulate} {Human} {Samples}.
\newblock \emph{Political Analysis}, 31(3): 337--351.

\bibitem[{Bail(2024)}]{bail_can_2024}
Bail, C.~A. 2024.
\newblock Can {Generative} {AI} improve social science?
\newblock \emph{Proceedings of the National Academy of Sciences}, 121(21): e2314021121.

\bibitem[{Barrie, Palmer, and Spirling(2024)}]{barrie_replication_2024}
Barrie, C.; Palmer, A.; and Spirling, A. 2024.
\newblock Replication for {Language} {Models}.

\bibitem[{Berndsen and Van Der~Pligt(2005)}]{berndsen_risks_2005}
Berndsen, M.; and Van Der~Pligt, J. 2005.
\newblock Risks of meat: the relative impact of cognitive, affective and moral concerns.
\newblock \emph{Appetite}, 44(2): 195--205.

\bibitem[{Bisbee et~al.(2023)Bisbee, Clinton, Dorff, Kenkel, and Larson}]{bisbee_synthetic_2023}
Bisbee, J.; Clinton, J.; Dorff, C.; Kenkel, B.; and Larson, J. 2023.
\newblock Synthetic {Replacements} for {Human} {Survey} {Data}? {The} {Perils} of {Large} {Language} {Models}.
\newblock preprint, SocArXiv.

\bibitem[{Bolderdijk and Cornelissen(2022)}]{bolderdijk_how_2022}
Bolderdijk, J.~W.; and Cornelissen, G. 2022.
\newblock “{How} do you know someone's vegan?” {They} won't always tell you. {An} empirical test of the do-gooder's dilemma.
\newblock \emph{Appetite}, 168: 105719.

\bibitem[{Carfora et~al.(2019)Carfora, Catellani, Caso, and Conner}]{carfora_how_2019}
Carfora, V.; Catellani, P.; Caso, D.; and Conner, M. 2019.
\newblock How to reduce red and processed meat consumption by daily text messages targeting environment or health benefits.
\newblock \emph{Journal of Environmental Psychology}, 65: 101319.

\bibitem[{Davidson et~al.(2024)Davidson, Veselovsky, Josifoski, Peyrard, Bosselut, Kosinski, and West}]{davidson_evaluating_2024}
Davidson, T.~R.; Veselovsky, V.; Josifoski, M.; Peyrard, M.; Bosselut, A.; Kosinski, M.; and West, R. 2024.
\newblock Evaluating {Language} {Model} {Agency} through {Negotiations}.
\newblock ArXiv:2401.04536 [cs].

\bibitem[{Dillard and Shen(2005)}]{dillard_nature_2005}
Dillard, J.~P.; and Shen, L. 2005.
\newblock On the {Nature} of {Reactance} and its {Role} in {Persuasive} {Health} {Communication}.
\newblock \emph{Communication Monographs}, 72(2): 144--168.

\bibitem[{Godfray et~al.(2018)Godfray, Aveyard, Garnett, Hall, Key, Lorimer, Pierrehumbert, Scarborough, Springmann, and Jebb}]{godfray_meat_2018}
Godfray, H. C.~J.; Aveyard, P.; Garnett, T.; Hall, J.~W.; Key, T.~J.; Lorimer, J.; Pierrehumbert, R.~T.; Scarborough, P.; Springmann, M.; and Jebb, S.~A. 2018.
\newblock Meat consumption, health, and the environment.
\newblock \emph{Science}, 361(6399): eaam5324.

\bibitem[{Graça, Calheiros, and Oliveira(2015)}]{graca_attached_2015}
Graça, J.; Calheiros, M.~M.; and Oliveira, A. 2015.
\newblock Attached to meat? ({Un}){Willingness} and intentions to adopt a more plant-based diet.
\newblock \emph{Appetite}, 95: 113--125.

\bibitem[{Harguess, Crespo, and Hong(2020)}]{harguess_strategies_2020}
Harguess, J.~M.; Crespo, N.~C.; and Hong, M.~Y. 2020.
\newblock Strategies to reduce meat consumption: {A} systematic literature review of experimental studies.
\newblock \emph{Appetite}, 144: 104478.

\bibitem[{Hewitt et~al.(2024)Hewitt, Ashokkumar, Ghezae, and Willer}]{hewitt_predicting_2024}
Hewitt, L.; Ashokkumar, A.; Ghezae, I.; and Willer, R. 2024.
\newblock Predicting {Results} of {Social} {Science} {Experiments} {Using} {Large} {Language} {Models}.

\bibitem[{Judge et~al.(2024)Judge, Bouman, Steg, and Bolderdijk}]{judge_accelerating_2024}
Judge, M.; Bouman, T.; Steg, L.; and Bolderdijk, J.~W. 2024.
\newblock Accelerating social tipping points in sustainable behaviors: {Insights} from a dynamic model of moralized social change.
\newblock \emph{One Earth}, 7(5): 759--770.

\bibitem[{Kwasny, Dobernig, and Riefler(2022)}]{kwasny_towards_2022}
Kwasny, T.; Dobernig, K.; and Riefler, P. 2022.
\newblock Towards reduced meat consumption: {A} systematic literature review of intervention effectiveness, 2001–2019.
\newblock \emph{Appetite}, 168: 105739.

\bibitem[{Monin, Sawyer, and Marquez(2008)}]{monin_rejection_2008}
Monin, B.; Sawyer, P.~J.; and Marquez, M.~J. 2008.
\newblock The rejection of moral rebels: {Resenting} those who do the right thing.
\newblock \emph{Journal of Personality and Social Psychology}, 95(1): 76--93.

\bibitem[{Pabian et~al.(2020)Pabian, Hudders, Poels, Stoffelen, and De~Backer}]{pabian_ninety_2020}
Pabian, S.; Hudders, L.; Poels, K.; Stoffelen, F.; and De~Backer, C. J.~S. 2020.
\newblock Ninety {Minutes} to {Reduce} {One}'s {Intention} to {Eat} {Meat}: {A} {Preliminary} {Experimental} {Investigation} on the {Effect} of {Watching} the {Cowspiracy} {Documentary} on {Intention} to {Reduce} {Meat} {Consumption}.
\newblock \emph{Frontiers in Communication}, 5: 69.

\bibitem[{Park et~al.(2022)Park, Popowski, Cai, Morris, Liang, and Bernstein}]{park_social_2022}
Park, J.~S.; Popowski, L.; Cai, C.; Morris, M.~R.; Liang, P.; and Bernstein, M.~S. 2022.
\newblock Social {Simulacra}: {Creating} {Populated} {Prototypes} for {Social} {Computing} {Systems}.
\newblock In \emph{Proceedings of the 35th {Annual} {ACM} {Symposium} on {User} {Interface} {Software} and {Technology}}, 1--18. Bend OR USA: ACM.
\newblock ISBN 978-1-4503-9320-1.

\bibitem[{Romo and Donovan-Kicken(2012)}]{romo_actually_2012}
Romo, L.~K.; and Donovan-Kicken, E. 2012.
\newblock “{Actually}, {I} {Don}'t {Eat} {Meat}”: {A} {Multiple}-{Goals} {Perspective} of {Communication} {About} {Vegetarianism}.
\newblock \emph{Communication Studies}, 63(4): 405--420.

\bibitem[{Saba and Di~Natale(1998)}]{saba_study_1998}
Saba, A.; and Di~Natale, R. 1998.
\newblock A study on the mediating role of intention in the impact of habit and attitude on meat consumption.
\newblock \emph{Food Quality and Preference}, 10(1): 69--77.

\bibitem[{Sen et~al.(2025)Sen, Lutz, Rogers, Garcia, and Strohmaier}]{sen_missing_2025}
Sen, I.; Lutz, M.; Rogers, E.; Garcia, D.; and Strohmaier, M. 2025.
\newblock Missing the {Margins}: {A} {Systematic} {Literature} {Review} on the {Demographic} {Representativeness} of {LLMs}.

\bibitem[{Sheeran et~al.(2005)Sheeran, Milne, Webb, and Gollwitzer}]{sheeran_implementation_2005}
Sheeran, P.; Milne, S.; Webb, T.~L.; and Gollwitzer, P.~M. 2005.
\newblock Implementation {Intentions} and {Health} {Behaviour}.
\newblock In Conner, M., ed., \emph{Predicting {Health} {Behaviour}: {Research} and {Practice} with {Social} {Cognition} {Models}}, 276--323. New York: Open Univ. Pr.
\newblock ISBN 978-0-335-21177-7.

\bibitem[{Shen, Sheer, and Li(2015)}]{shen_impact_2015}
Shen, F.; Sheer, V.~C.; and Li, R. 2015.
\newblock Impact of {Narratives} on {Persuasion} in {Health} {Communication}: {A} {Meta}-{Analysis}.
\newblock \emph{Journal of Advertising}, 44(2): 105--113.

\bibitem[{Taubenfeld et~al.(2024)Taubenfeld, Dover, Reichart, and Goldstein}]{taubenfeld_systematic_2024}
Taubenfeld, A.; Dover, Y.; Reichart, R.; and Goldstein, A. 2024.
\newblock Systematic {Biases} in {LLM} {Simulations} of {Debates}.
\newblock In \emph{Proceedings of the 2024 {Conference} on {Empirical} {Methods} in {Natural} {Language} {Processing}}, 251--267. Miami, Florida, USA: Association for Computational Linguistics.

\bibitem[{Thürmer, Stadler, and McCrea(2022)}]{thurmer_intergroup_2022}
Thürmer, J.~L.; Stadler, J.; and McCrea, S.~M. 2022.
\newblock Intergroup {Sensitivity} and {Promoting} {Sustainable} {Consumption}: {Meat} {Eaters} {Reject} {Vegans}’ {Call} for a {Plant}-{Based} {Diet}.
\newblock \emph{Sustainability}, 14(3): 1741.

\bibitem[{Tjuatja et~al.(2024)Tjuatja, Chen, Wu, Talwalkar, and Neubig}]{tjuatja_llms_2024}
Tjuatja, L.; Chen, V.; Wu, S.~T.; Talwalkar, A.; and Neubig, G. 2024.
\newblock Do {LLMs} exhibit human-like response biases? {A} case study in survey design.
\newblock ArXiv:2311.04076 [cs].

\bibitem[{Törnberg et~al.(2023)Törnberg, Valeeva, Uitermark, and Bail}]{tornberg_simulating_2023}
Törnberg, P.; Valeeva, D.; Uitermark, J.; and Bail, C. 2023.
\newblock Simulating {Social} {Media} {Using} {Large} {Language} {Models} to {Evaluate} {Alternative} {News} {Feed} {Algorithms}.
\newblock ArXiv:2310.05984 [cs].

\bibitem[{Vaccaro et~al.(2025)Vaccaro, Caoson, Ju, Aral, and Curhan}]{vaccaro_advancing_2025}
Vaccaro, M.; Caoson, M.; Ju, H.; Aral, S.; and Curhan, J.~R. 2025.
\newblock Advancing {AI} {Negotiations}: {New} {Theory} and {Evidence} from a {Large}-{Scale} {Autonomous} {Negotiations} {Competition}.
\newblock Version Number: 1.

\bibitem[{Wang, Morgenstern, and Dickerson(2024)}]{wang_large_2024}
Wang, A.; Morgenstern, J.; and Dickerson, J.~P. 2024.
\newblock Large language models cannot replace human participants because they cannot portray identity groups.
\newblock ArXiv:2402.01908 [cs].

\bibitem[{Wyker and Davison(2010)}]{wyker_behavioral_2010}
Wyker, B.~A.; and Davison, K.~K. 2010.
\newblock Behavioral {Change} {Theories} {Can} {Inform} the {Prediction} of {Young} {Adults}' {Adoption} of a {Plant}-based {Diet}.
\newblock \emph{Journal of Nutrition Education and Behavior}, 42(3): 168--177.

\end{thebibliography}

\appendix

\section{Questionnaires} \label{app_sec:questionnaire}

In the following section, the questionnaires used in this study to assess all measured outcomes are described. An overview of the six key constructs is provided in Figure~\ref{fig:tpb}.

To assess the \textit{persuasiveness} of the presented arguments, participants were asked to rate the item \textit{"How persuasive did you find this argument?"} on a 7-point Likert scale.

\textit{Behavioral change} was measured with the item \textit{"To what extent did this argument make you consider changing your behavior?"}, also using a 7-point Likert scale.

\textit{Attitudes toward meat reduction} were assessed using a four-item semantic differential scale based on \citet{berndsen_risks_2005}, rated on a 7-point scale:
\begin{itemize}
  \item Pleasant -- Unpleasant
  \item Useful -- Useless
  \item Favorable -- Unfavorable
  \item Good -- Bad
\end{itemize}

\textit{Intentions to reduce meat consumption} were captured with a three-item scale adapted from \citet{graca_attached_2015}, using a 7-point Likert scale:
\begin{itemize}
  \item I intend to eat less white meat.
  \item I intend to eat less red meat.
  \item I intend to eat less processed meat.
\end{itemize}

\textit{Subjective norms} were measured with four items adapted from \citet{berndsen_risks_2005}, each rated on a 7-point scale:
\begin{itemize}
  \item How much do you feel your friends want you to reduce your meat consumption in the next year?
  \item How much do you feel your family want you to reduce your meat consumption in the next year?
  \item How much do you feel health experts want you to reduce your meat consumption in the next year?
  \item How much do you feel your colleagues want you to reduce your meat consumption in the next year?
\end{itemize}

\textit{Perceived behavioral control} was assessed via two items based on \citet{wyker_behavioral_2010}, using a 7-point Likert scale:
\begin{itemize}
  \item How much personal control do you feel you have about reducing your meat consumption in the next year?
  \item To what extent do you see yourself as being capable of reducing your meat consumption in the next year?
\end{itemize}

\textit{Attachment to meat} was measured using the 20-item Meat Attachment Questionnaire (MAQ) developed by \citet{graca_attached_2015}, on a 5-point Likert scale. Items include:
\begin{itemize}
  \item To eat meat is one of the good pleasures in life.
  \item Meat is irreplaceable in my diet.
  \item According to our position in the food chain, we have the right to eat meat.
  \item I feel bad when I think of eating meat.
  \item I love meals with meat.
  \item To eat meat is disrespectful towards life and the environment.
  \item To eat meat is an unquestionable right of every person.
  \item Meat consumption is crucial to my balance.
  \item A full meal is a meal with meat.
  \item I'm a big fan of meat.
  \item If I couldn't eat meat I would feel weak.
  \item If I was forced to stop eating meat I would feel sad.
  \item By eating meat I'm reminded of the death and suffering of animals.
  \item Eating meat is a natural and undisputable practice.
  \item I don't picture myself without eating meat regularly.
  \item Meat sickens me.
  \item I would feel fine with a meatless diet.
  \item Meat consumption is a natural act of one's affirmation as a human being.
  \item A good steak is without comparison.
  \item Meat reminds me of diseases.
\end{itemize}

The \textit{perceived threat to freedom} was assessed using four items from \citet{dillard_nature_2005}, rated on a 7-point Likert scale:
\begin{itemize}
  \item The person tried to make a decision about my own diet for me.
  \item The person tried to influence me and my diet.
  \item The person threatened my freedom to decide about my own diet.
  \item The person tried to pressure me regarding my diet.
\end{itemize}

\textit{Social attractiveness} was measured with seven items adapted from \citet{thurmer_intergroup_2022}, using a 7-point Likert scale:
\begin{itemize}
  \item To what extent do you think your interaction partner is intelligent?
  \item ...trustworthy?
  \item ...friendly?
  \item ...open?
  \item ...likeable?
  \item ...respectable?
  \item ...interesting?
\end{itemize}

\textit{Social closeness} was measured via seven items based on \citet{monin_rejection_2008}, rated on a 7-point Likert scale:
\begin{itemize}
  \item I would like to get to know the other person better.
  \item I would like to have lunch with the other person sometime.
  \item I would like to work together with the other person on a task.
  \item I would like to have a conversation with the other person about food.
  \item I felt emotionally close to the other person.
  \item I wanted to have a conversation with the other person.
  \item I felt like I made friends with the other person.
\end{itemize}

\section{Additional Results} \label{app_sec:additional_results}

\begin{figure*}[ht!]
    \centering
    \hfill
    \begin{subfigure}[b]{0.33\linewidth}
        \centering
        \includegraphics[width=\linewidth]{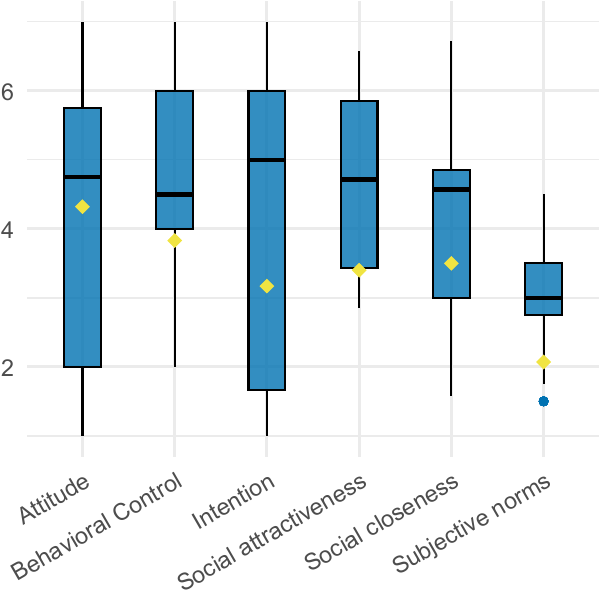}
        \subcaption{\textbf{Llama3.3 70B}}
        \label{app_fig:boxplot_70B}
    \end{subfigure}
    \hfill
    \begin{subfigure}[b]{0.33\linewidth}
        \centering
        \includegraphics[width=\linewidth]{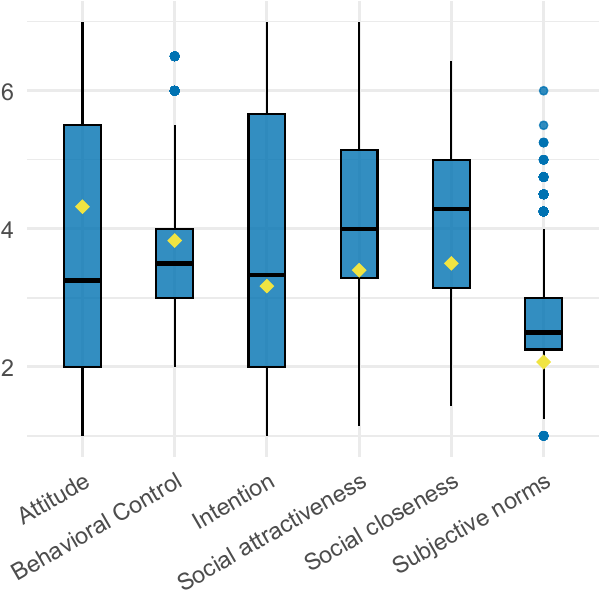}
        \subcaption{\textbf{Llama 3.1 8B}}
        \label{app_fig:boxplot_8B}
    \end{subfigure}
    \hfill
    \begin{subfigure}[b]{0.33\linewidth}
        \centering
        \includegraphics[width=\linewidth]{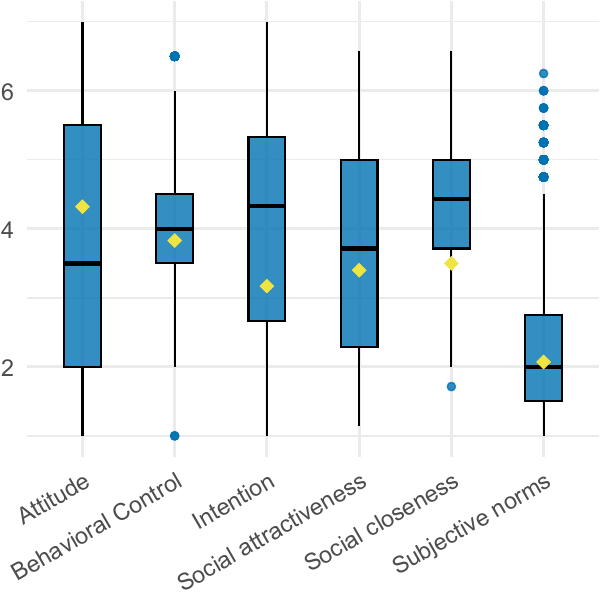}
        \subcaption{\textbf{Llama 3.2 3B}}
        \label{app_fig:boxplot_3B}
    \end{subfigure}
    \hfill
    \caption{\textbf{Smaller \texttt{Llama} models show less consistent construct response distributions.} The distribution of model-generated responses is shown across \texttt{Llama models (70B, 8B, 3B)} for key psychological constructs. Yellow diamonds indicate the mean from human samples for each construct. Compared to the {\texttt70B} model, smaller models exhibit reduced variability and more outliers in several constructs (e.g., \textit{Behavioral Control}, \textit{Subjective Norms}), suggesting a loss of distributional alignment with human responses as model size decreases.}
    \label{app_fig:boxplots}
\end{figure*}

\begin{table*}[ht!]
    \centering
    \footnotesize
    \newcolumntype{L}[1]{>{\raggedright\arraybackslash}p{#1}}
    \newcolumntype{R}[1]{>{\raggedleft\arraybackslash}p{#1}}
    \begin{tabular}{l R{1cm} | R{1cm} R{1cm} R{1cm} | R{1.2cm} l}
         \textbf{Scale} & \textbf{\#items} & \raggedright\textbf{Llama 3.3 70B} & \raggedright\textbf{Llama 3.1 8B} & \raggedright\textbf{Llama 3.2 3B} & \raggedright\textbf{Human Samples} \ & \textbf{Original Study} \\
         \toprule
         \textbf{Theory of Planned Behavior} & & & & & & \\
         Attitude              & 4 & 
         .99 & .97 & .92 & .86 & \citet{pabian_ninety_2020}\\
         Intention             & 3 & 
         .99 & .93 & .88 & .85 & \citet{pabian_ninety_2020}\\
         Subjective Norms      & 2 & 
         .76 & \textcolor{okabe_red}{.59}& .78 & .75 & \citet{wyker_behavioral_2010}\\
         Behavioral Control    & 2 & 
         .70 & \textcolor{okabe_red}{.66} & \textcolor{okabe_red}{.68} & .70 & \citet{wyker_behavioral_2010}\\
         \midrule
         \textbf{Social Costs} & & & & & & \\
         Social Attractiveness & 7 & 
         .98 & .93 & .96 & .91 & \citet{thurmer_intergroup_2022} \\
         Social Closeness      & 7 & 
         .98 & .94 & .82 & .91 & \citet{monin_rejection_2008}\\
         \midrule
         \textbf{Additional Constructs} & & & & & & \\
         Threat of Freedom     & 4 & 
         .93 & .87 & \textcolor{okabe_red}{.60} & .84 & \citet{dillard_nature_2005}\\
         Meat Attachment       & 20 & .99 & .98 & .71 & .92 & \citet{graca_attached_2015}\\
    \end{tabular}
    \caption{\textbf{Reliability of psychological constructs decreases with smaller Llama models.} Cronbach’s Alpha values for each scale are compared across \texttt{Llama models (3.3 70B, 3.1 8B, 3.2 3B)} and human participant data to assess internal consistency. While larger models consistently show high reliability ($\alpha \geq .70$), several scales fall below the acceptable threshold in smaller models (shown in \textcolor{okabe_red}{red})---most notably \textit{Behavioral Control}, and \textit{Threat of Freedom}. This suggests a degradation in the coherent representation of multi-item constructs as model size decreases.}
    \label{app_tab:cronbachs_alpha}
\end{table*}

\begin{figure*}[ht!]
    \centering
    \hfill
    \begin{subfigure}[b]{0.31\linewidth}
        \centering
        \includegraphics[width=\linewidth]{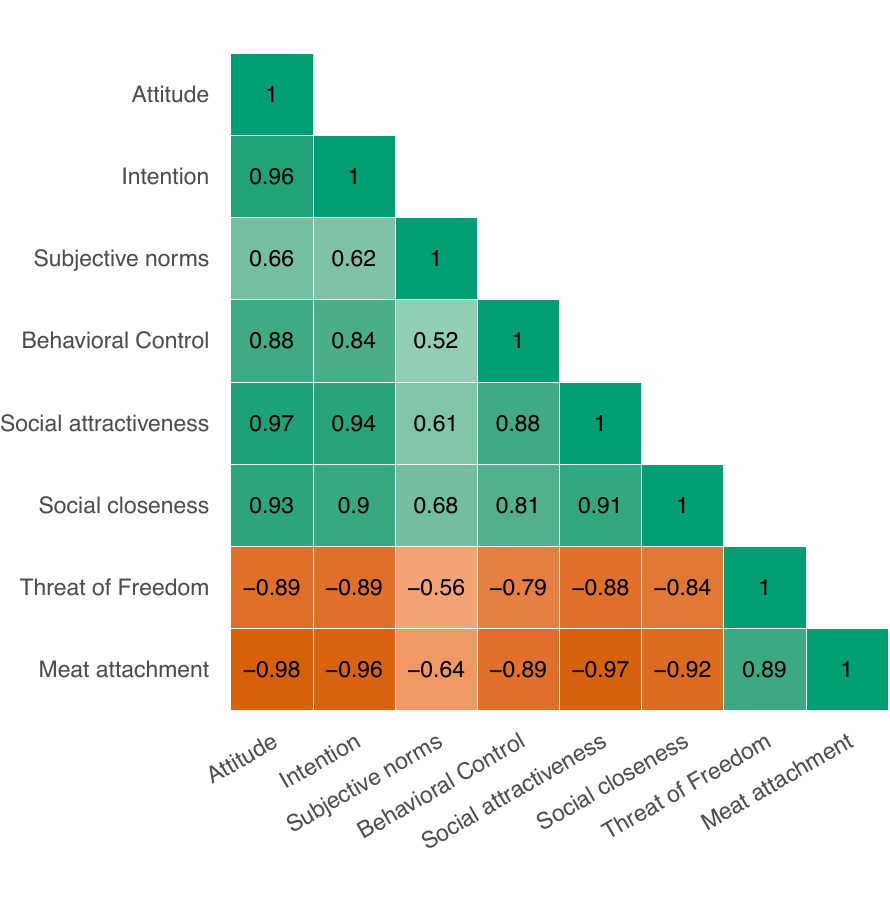}
        \subcaption{\textbf{Llama 3.3 70B}}
        \label{app_fig:heatmap_70B}
    \end{subfigure}
    \hfill
    \begin{subfigure}[b]{0.31\linewidth}
        \centering
        \includegraphics[width=\linewidth]{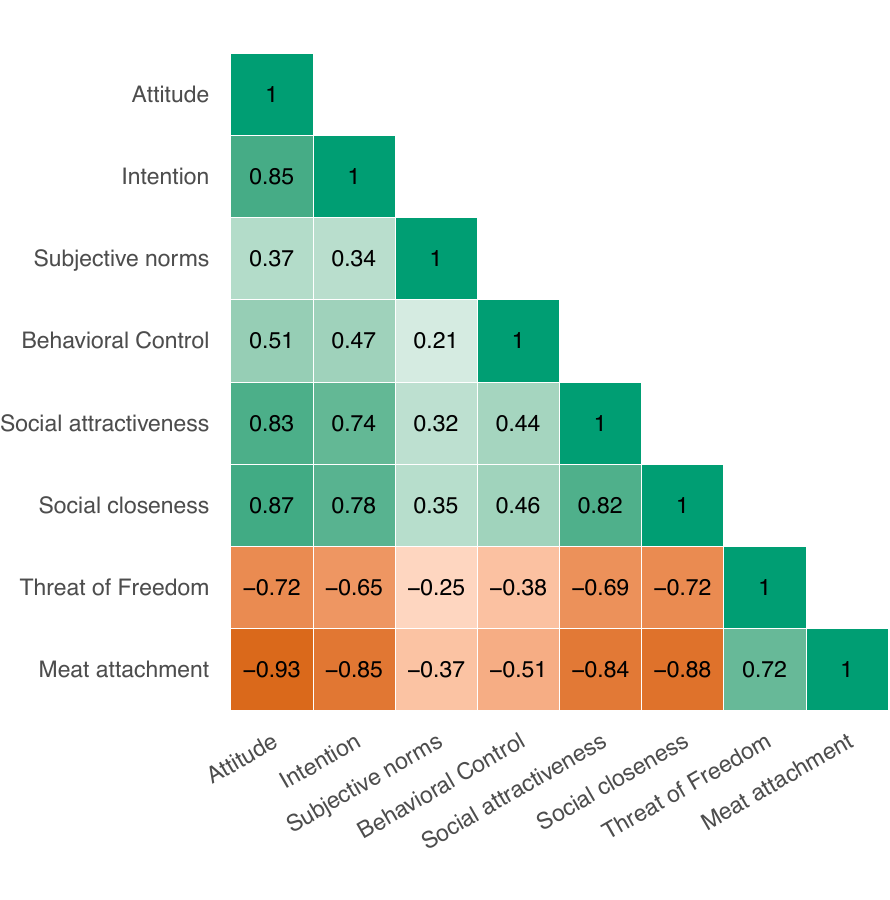}
        \subcaption{\textbf{Llama 3.1 8B}}
        \label{app_fig:heatmap_8B}
    \end{subfigure}
    \hfill
    \begin{subfigure}[b]{0.365\linewidth}
        \centering
        \includegraphics[width=\linewidth]{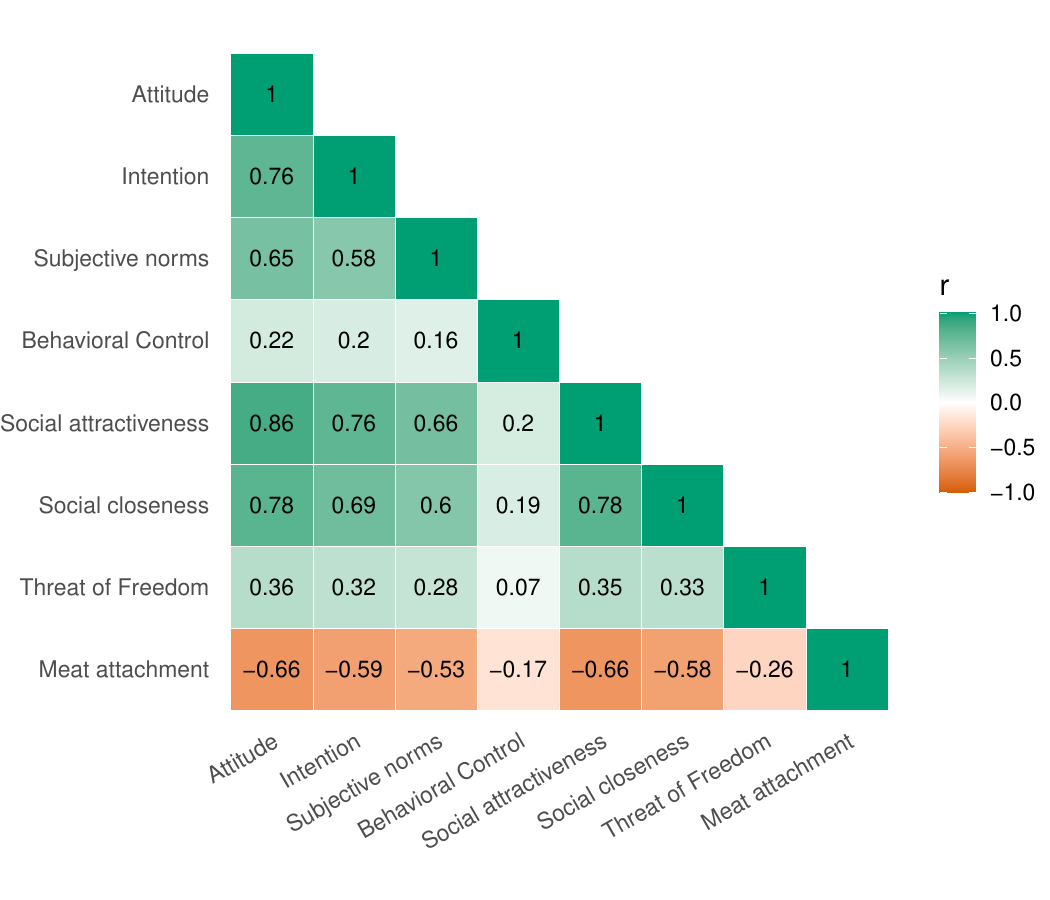}
        \subcaption{\textbf{Llama 3.2 3B}}
        \label{app_fig:heatmap_3B}
    \end{subfigure}
    \hfill
    \caption{\textbf{Correlation strength weakens with smaller Llama models, while overall construct patterns remain stable.} To assess construct validity, we examine Pearson correlations among psychological constructs across three Llama models of varying sizes. While general correlation patterns---including the two added constructs \textit{Threat of Freedom} and \textit{Meat Attachment}---remain largely stable, the strength of correlations systematically decreases as model size reduces. This suggests a loss of representational consistency in smaller models.}
    \label{app_fig:heatmaps}
\end{figure*}

\begin{figure*}[ht!]
    \centering
    \hfill
    \begin{subfigure}[b]{0.392\linewidth}
        \centering
        \includegraphics[width=\linewidth]{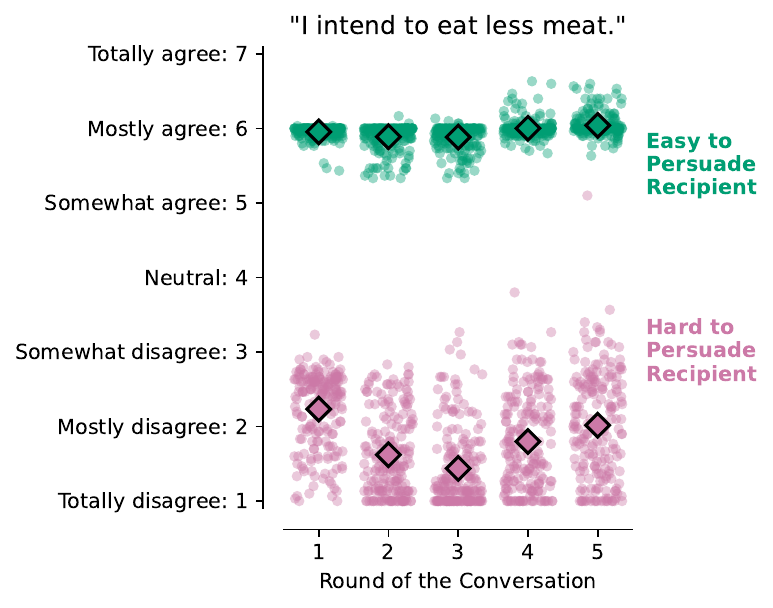}
        \subcaption{\textbf{Llama 3.3 70B}}
        \label{app_fig:meat_intention_70B}
    \end{subfigure}
    \hfill
    \begin{subfigure}[b]{0.28\linewidth}
        \centering
        \includegraphics[width=\linewidth]{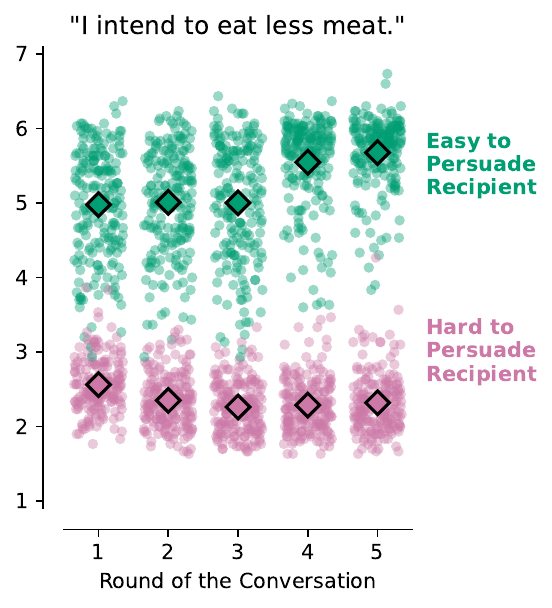}
        \subcaption{\textbf{Llama 3.1 8B}}
        \label{app_fig:meat_intention_8B}
    \end{subfigure}
    \hfill
    \begin{subfigure}[b]{0.28\linewidth}
        \centering
        \includegraphics[width=\linewidth]{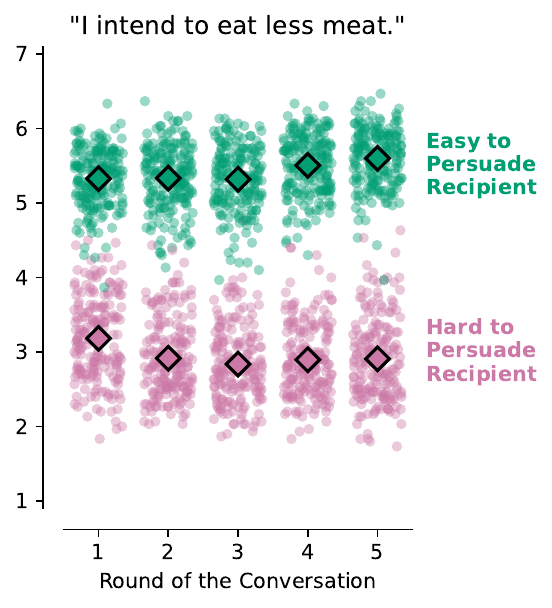}
        \subcaption{\textbf{Llama 3.2 3B}}
        \label{app_fig:meat_intention_3B}
    \end{subfigure}
    \hfill
    \caption{\textbf{Recipients’ Response to: “I Intend to Eat Less Meat”}, obtained from simulations run with \texttt{Llama 3.3 70B}, \texttt{Llama 3.1 8B} and \texttt{Llama 3.2 3B}. Combined survey responses after each conversation round, where $\Diamond$ indicates the mean over all conversations per round. Compared to \texttt{Llama 3.3 70B}, we observe much greater variance in survey responses for simulations performed with smaller models. The intention of the \textcolor{okabe_purple}{Hard to Persuade Recipient} stays more stable over time for the smaller models, while the intention of the \textcolor{okabe_green}{Easy to Persuade Recipient} shows a stronger increase in round 4, especially for \texttt{Llama 3.1 8B}.}
    \label{app_fig:meat_intention}
\end{figure*}

%\begin{table*}[ht!]
%    \centering
%    \begin{tabular}{c|c}
%         &  \\
%         & 
%    \end{tabular}
%    \caption{\textbf{Most and Least Successful Persuasion Strategies}}
%    \label{app_tab:strategies}
%\end{table*}

\end{document}